\documentstyle[aps,amssymb,eqsecnum,preprint,tighten,epsf]{revtex}
\draft

  \newcommand{\bcone}{{\bf BA2}}
  \newcommand{\bctwo}{{\bf BA3}}
  \newcommand{\bcthree}{{\bf BA5}}
  \newcommand{\bcfour}{{\bf BA4}}
  \newcommand{\bcfive}{{\bf BA1}}

\begin{document}

\title{\rightline{\small{{\it Phys. Rev. D} {\bf 62}, 104006 (2000)\/}}
Cauchy boundaries in linearized gravitational theory}

\author{
	Bela Szil\'{a}gyi${}^{1}$ ,
        Roberto G\'omez${}^{1}$,
        Nigel T. Bishop${}^{2}$ and
	Jeffrey Winicour${}^{1,3}$
       }
\address{
${}^{1}$ Department of Physics and Astronomy \\
         University of Pittsburgh, Pittsburgh, PA 15260, USA \\
${}^{2}$ Department of Mathematics, Applied Mathematics and Astronomy\\
         University of South Africa,
         P.O. Box 392, Pretoria 0003, South Africa \\
${}^{3}$ Max-Planck-Institut f\" ur
         Gravitationsphysik, Albert-Einstein-Institut, \\
	 14476 Golm, Germany
	 }
\date{\today}
\maketitle

\begin{abstract}

We investigate the numerical stability of Cauchy evolution of
linearized gravitational theory in a 3-dimensional bounded domain.
Criteria of robust stability are proposed, developed into a testbed and
used to study various evolution-boundary algorithms.  We construct a
standard explicit finite difference code which solves the unconstrained
linearized Einstein equations in the $3+1$ formulation and measure its
stability properties under Dirichlet, Neumann and Sommerfeld boundary
conditions. We demonstrate the robust stability of a specific
evolution-boundary algorithm under random constraint violating initial
data and random boundary data.

\end{abstract}

\pacs{PACS number(s): 04.20Ex, 04.25Dm, 04.25Nx, 04.70Bw}

\section{Introduction}

The computational evolution of 3-dimensional general relativistic
space-times by means of Cauchy evolution is a potentially powerful tool
to study the gravitational radiation from black-hole/neutron-star
binaries whose inspiral are expected to provide prominent signals to
gravitational wave observatories.  There are several 3-dimensional
general relativistic codes under development to solve this problem.
Boundary conditions are an essential part of these codes.  At the outer
boundary they must provide an outgoing radiation condition and extract
the emitted waveform. For black-hole spacetimes, there is also an inner
boundary, approximately given by the apparent horizon, where one
excises the singular region inside a black hole. Instabilities or
inaccuracies introduced at such boundaries have emerged as a major
problem common to all code development. Historically, the first Cauchy
codes were based upon the Arnowitt-Deser-Misner (ADM)
formulation~\cite{adm,york} of the Einstein equations. Recently there
has been pessimism that such codes might be inherently unstable because
of the lack of manifest hyperbolicity in the underlying equations. In
order to shed light on this issue, we present here a study of ADM
evolution-boundary algorithms in the simple environment of
linearized gravity, where nonlinear sources of physical or numerical
instability are absent and computing time is reduced by a factor of
five by use of a linearized code.

Our two main results, for the case of fixed lapse and shift, are:

\begin{itemize}
\item  On analytic grounds, ADM boundary algorithms which supply values
for all components of the metric (or extrinsic curvature) are
inconsistent.

\item We present a boundary algorithm which allows free specification of
the transverse-traceless components of the metric (or extrinsic
curvature) at the boundary, and for which unconstrained, linearized ADM
evolution can be carried out in a bounded domain for thousands of
crossing times with robust stability.
\end{itemize}

The criteria for robust stability, which we present here, are the most
severe that have been applied to Cauchy evolution in numerical relativity.
The boundary algorithm differs from previous approaches and offers
fresh hope for robust nonlinear ADM evolution.

Our particular motivation for this work is the difficulty we have experienced
implementing Cauchy-characteristic-matching (CCM) for 3-dimensional general
relativity~\cite{vishu,cqg}. CCM provides a Cauchy boundary condition by
matching the Cauchy evolution across the boundary to a characteristic
evolution. For nonlinear scalar waves propagating in a flat 3-dimensional
space, CCM has been demonstrated to be more accurate and efficient than all
other existing boundary conditions for Cauchy evolution~\cite{ccm}, and it has
been demonstrated mathematically that this conclusion also applies to
gravity~\cite{cce}. In addition, in the spherically symmetric case of a self
gravitating scalar wave satisfying the Einstein-Klein-Gordon equations, CCM has
been successfully applied at the inner boundary of a Cauchy evolution to excise
the interior black hole region and, at the same time, at the outer boundary to
provide a global evolution on a compactified  grid extending to null
infinity~\cite{marsa}. These successes are promising for the application of CCM
to 3-dimensional problems in general relativity but this has not yet been borne
out. This difficulty, and the similar difficulty in efforts using perturbative
matching~\cite{rezz}, may possibly arise from a pathology of the Cauchy
boundary which is independent of matching. In this work, we reveal such a
pathology in the way boundary conditions have been applied in the ADM
formulation of the Einstein equations which, at present, is the only
formulation for which matching has been attempted. We also present a new form
of ADM boundary algorithm which eliminates the pathology.

The stability of the Cauchy evolution algorithm itself is
straightforward to investigate by carrying out a boundary-free
evolution on a 3-torus (equivalent to periodic boundary conditions).
Such tests constitute Stage 1 of a 3-stage test bed for robust boundary
stability which is summarized below and explained in detail in
Sections~\ref{sec:boun},~\ref{sec:stage2} and ~\ref{sec:stage3}.  The
periodic boundary tests serve to cull out algorithms whose boundary
stability is doomed from the start. In earlier work, robust stability
for characteristic evolution with random data on an inner boundary was
demonstrated for characteristic evolution using the PITT Null
Code~\cite{cce}. In the course of the present investigation we have
reconfirmed this robustness of the PITT code using the same
specifications proposed here for Cauchy codes.

CCM cannot work unless the Cauchy code, as well
as the characteristic code, has a robustly stable boundary. This is
necessarily so because the interpolations between a Cartesian Cauchy
grid and a spherical null grid  continually introduce short wavelength
noise into the neighborhood of the boundary. This is the rationale
underlying the robustness criterion in our test bed. Robustness of the
Cauchy boundary is a necessary (although not a sufficient) condition
for the successful implementation of CCM.

Analytic studies of Cauchy evolution of linearized gravity with
boundaries at infinity reveal modes which grow linearly in time, but
none which grow exponentially~\cite{matzn}. The inaccuracy introduced
by such secular modes can be controlled and is not of major concern, at
least in the linearized theory.  (Such secular modes can lead to
exponential instabilities of numerical origin in the nonlinear theory
if not properly treated~\cite{aabss}). In the case of a finite
boundary, there is further potential for instability and a brief
discussion is given in Sec.~\ref{sec:boun}.

As is customary in numerical relativity, we monitor the existence of unstable
modes by the growth of the Hamiltonian constraint. Because the constraints are
not enforced during standard implementation of ADM evolution, the Hamiltonian
constraint is an effective sensor of numerical instabilities.

Stage 2 of the test bed is based on the simple boundary value problem obtained
by opening one dimension of a 3-torus to form a 2-torus with plane boundaries
normal to a Cartesian axis. Running a Cauchy-boundary algorithm with this
topology and with random initial and random boundary data forms the second
stage of our test bed, which is discussed in Sec.~\ref{sec:stage2}. In
Sec.~\ref{sec:st2bc}, we present new evolution-boundary algorithms
which are robustly stable.

Stage 3 of the testbed is designed to test robustness of boundary
conditions appropriate to an isolated system. In Sec.~\ref{sec:stage3} we
establish Stage 3 robustness of an ADM boundary algorithm.

The main results presented here are experimental, in a computational
sense. The difficulties encountered with finite Cauchy boundaries in
general relativity have recently prompted some analytic investigations
of the subject~\cite{stewart,nagy}. However, these have so far been
confined to hyperbolic formulations, as opposed to the ADM formulation,
and to the analytic problem, as opposed to the finite difference
solution obtained by computation.  Although it is not possible to make
a direct comparison, the nature of our results are consistent with
the general conclusions of these analytic studies.

There are several promising numerical approaches based upon hyperbolic (or
``more hyperbolic'') formulations of the
equations~\cite{bmasso,newt,shibnak,fried96,fritreul,abrahams97,scheel97,bmss,baumgsh,robmasso,phub}.
Here we concentrate on ADM schemes, which are the most compact to code and
require the least amount of memory because they have a smaller number of
variables. Our results should provide useful benchmarks for other relativity
codes. However, it should also be cautioned that the nature of a successful
boundary algorithm is dependent on the form of the equations adopted, as well
as the choice of discretization, and the ADM boundary algorithms we have
obtained do not necessarily apply to other formulations.

We use Greek letters for space-time indices and Latin letters for spatial
indices. Four dimensional geometric quantities are explicitly indicated,
such as ${}^{(4)}R_{\alpha\beta}$ and ${}^{(4)}G_{\alpha\beta}$ for the
Ricci and Einstein tensors of the space-time, whereas $R_{ij}$ and $R$
refer to the Ricci tensor and Ricci scalar of the Cauchy hypersurfaces.
Linearized versions of these quantities are denoted by ${}^{(4)}\tilde
R_{\alpha\beta}$, $\tilde R_{ij}$, etc. Three dimensional tensor indices
are raised and lowered by the background Euclidean metric $\delta_{ij}$.
We write $h=\delta^{ij}h_{ij}$ for 3-dimensional traces. We denote time
derivatives by $\dot f= \partial_t f$. Our convention for the background
Minkowski metric is such that the wave equation takes the form
\begin{equation}
     \eta^{\alpha\beta}\partial_ \alpha \partial_\beta \Phi
    =(- \partial_t^2 + \partial_x^2  + \partial_y^2 +\partial_z^2 )\Phi
         =0.
\label{eq:wave}
\end{equation}

\section{General Framework}

\subsection{The linearized ADM system} \label{sec:admsystem}

The ADM formulation of the Einstein equations  introduces a foliation of
space-time by a time coordinate $t$ and expresses the 4-dimensional
metric as
\begin{equation}
  ds^2 = -\alpha^2 dt^2 + g_{ij} \left(dx^i + \beta^i dt\right)
                                 \left(dx^j + \beta^j dt\right) ,
\end{equation}
where $g_{ij}$ is the induced 3-metric of the $t=const$ slices,
$\alpha$ is the lapse and $\beta^i$ the shift, with the normal
to the slices given by $n^\mu=(1,-\beta^i)/\alpha$.  The equations
${}^{(4)}R_{ij}=0$ yield the evolution equations
\begin{eqnarray}
   \partial_t g_{ij} -\pounds_\beta g_{ij} &=& -2\alpha K_{ij} \\
   \partial_t K_{ij} -\pounds_\beta K_{ij} &=& -D_i D_j\alpha
    + \alpha\left(R_{ij} + K K_{ij} - 2 K_i^l K_{lj} \right) ,
\end{eqnarray}
for the 3-metric $g_{ij}$ and the extrinsic curvature $K_{ij} =
-\frac{1}{2} \pounds_n g_{ij}$, subject to the constraints
\begin{eqnarray}
   R - K_{ij} K^{ij} + K^2 &=& 0 \\
   D_j \left( K^{ij} - g^{ij} K\right) &=& 0.
\end{eqnarray}
Here $R$, $R_{ij}$ and $D_i$ are the Ricci scalar, Ricci tensor and
connection of the 3-metric, respectively.

For simplicity we consider a gauge in which the lapse is unity and the
shift vanishes (Gaussian coordinates), so that the linearized metric
$g_{\alpha\beta}=\eta_{\alpha\beta}+h_{\alpha\beta}$  satisfies
$h_{t\alpha}=0$, and obeys the linearized ADM evolution equations
\begin{eqnarray}
   \partial_t h_{ij} &=& -2 K_{ij} \nonumber \\
   \partial_t K_{ij} &=& \tilde{R}_{ij} ,
\label{eq:linear}
\end{eqnarray}
subject to the (linearized) constraints
\begin{eqnarray}
   \tilde{R} &=& 0 \nonumber \\
   \partial_j \left( K^{ij} - \delta^{ij} K\right) &=& 0 .
\end{eqnarray}

Here we consider a 1-parameter system of equations, equivalent to the
linearized Einstein equation, consisting of the six evolution equations
${\cal E}_{ij}=0$ along with the four constraint equations ${\cal
C}={\cal C}_i=0$, where
\begin{equation}
  {\cal E}_{ij}:={}^{(4)}\tilde R_{ij}+\frac{1}{2}\lambda \delta_{ij} {\cal C},
\label{eq:evol}
\end{equation}
${\cal C}:={}^{(4)}\tilde G_{tt  }$, ${\cal C}_i:=-{}^{(4)}\tilde
G_{ti}$ and the parameter $\lambda$ allows mixing the
(linearized) Hamiltonian constraint ${\cal C}$ into the evolution
equations. For $\lambda=0$ we recover the standard ADM system.

Codes under development for the evolution of 3-dimensional space-times
without symmetry apply the constraint equations at the initial time but
do not enforce them during the evolution. It is crucial for this approach
that the constraints be stably propagated in time. An investigation by
Frittelli~\cite{frit} shows that this requires the parameter $\lambda$ in
Eq.~(\ref{eq:evol}) satisfy $1+\lambda \ge 0$. This follows from an
analysis of the linearized Bianchi identities $\partial_\beta
{}^{(4)}\tilde G_\alpha^\beta\equiv 0$, which imply that
\begin{eqnarray}
   \dot {\cal C}^i +  (1 + \lambda)\,
   \partial^i {\cal C} + \partial_j {\cal E}^{ij} \equiv 0 \\ \dot {\cal C}  +
   \partial_i {\cal C}^i \equiv 0.
\end{eqnarray}
Thus if the evolution equations are satisfied then the Hamiltonian
constraint satisfies
\begin{equation}
\ddot {\cal C} - (1 + \lambda) \partial^k \partial_k  {\cal
C}= 0; \label{eq:hwave}
\end{equation}
This equation has a well-posed initial value problem for $\lambda > -1$
(when it is hyperbolic) and also for $\lambda = -1$, but for $\lambda <
-1$ the equation is elliptic and the initial value problem is not
well-posed. In the standard ADM case, $\lambda=0$ and the
Hamiltonian constraint propagates along the light cone. We consider
here evolution equations with a range of $\lambda$.

The linearized evolution equations (\ref{eq:evol}) take the form
\begin{eqnarray}
   \dot h_{ij} &=& -2 K_{ij} \nonumber \\
   \dot K_{ij} &=& -\frac{1}{2} \partial_m \partial^m h_{ij}
             + \frac{1}{2}(\partial_i H_j +\partial_j H_i)
              +\frac{1}{2}\delta_{ij}\lambda {\cal C},
\label{eq:ewave}
\end{eqnarray}
where
\begin{equation}
       H_i =\partial^j(h_{ij}-\frac{1}{2} \delta_{ij}h).
\label{eq:Hi}
\end{equation}
and we can express the Hamiltonian as
\begin{equation}
      {\cal C}= \frac{1}{2} \partial_i H^i
              -\frac{1}{4}\partial_m\partial^m h.
\end{equation}
A spectral analysis of a system similar to Eq's (\ref{eq:ewave}) -
(\ref{eq:Hi}) is presented in~\cite{abdfpsst}.

\subsection{Finite difference algorithms}

The evolution variables consist of the 3-metric perturbations $h_{ij}$ and
their associated momentum $K_{ij}=-\dot h_{ij}/2$. The evolution is
implemented on a uniform spatial grid $(x_j,y_k,z_l) = (j\Delta x,k\Delta
x,l\Delta x)$  with time levels $t^n =n\Delta t$. The three different
evolution algorithms we apply can be discussed in reference to the scalar wave
Eq.~(\ref{eq:wave}), rewritten in the form
\begin{eqnarray}
   \dot \Phi &=& -2 K  \nonumber \\
   \dot K &=& -\frac{1}{2} \partial_m \partial^m \Phi,
\label{eq:canwave}
\end{eqnarray}
analogous to the first differential order in time and second differential order
in space form of the ADM equations. We denote $\Phi^n_{j,k,l}=\Phi(t^n,j\Delta
x,k\Delta x,l\Delta x)$. All second derivatives on the right hand side of
Eq.~(\ref{eq:canwave}) are calculated as centered 3-point finite differences.

\subsubsection{Standard leapfrog ({\bf LF1})}

The first evolution  algorithm, which we refer to as {\bf LF1}, is a standard
leapfrog implementation of Eq.~(\ref{eq:canwave}):
\begin{eqnarray}
      \Phi^{n+1}_{j,k,l} &=&\Phi^{n-1}_{j,k,l} -4 K^n_{j,k,l} \Delta t
                   \nonumber \\
   K^{n+1}_{j,k,l} &=& K^{n-1}_{j,k,l} -\nabla^2 \Phi^n_{j,k,l} \Delta t,
\end{eqnarray}
where $\nabla^2$ is the second order accurate centered difference
approximation to the Laplacian. It is known that this algorithm has a
time-splitting instability in the presence of dissipative and nonlinear
effects~\cite{cent}.

\subsubsection{Staggered leapfrog ({\bf LF2})}

The second evolution algorithm, which we refer to as {\bf LF2}, is a staggered
in time leapfrog scheme which is not subject to the time-splitting instability:
\begin{eqnarray}
   \Phi^{n+1}_{j,k,l} &=&\Phi^{n}_{j,k,l} -2 K^{n+\frac{1}{2}}_{j,k,l}\Delta t
\label{eq:lf21}\\
   K^{n+\frac{1}{2}}_{j,k,l}&=& K^{n-\frac{1}{2}}_{j,k,l} - \frac{1}{2}
         \nabla^2 \Phi^n_{j,k,l} \Delta t.
\label{eq:lf22}
\end{eqnarray}
Here $K$ is evaluated on the half grid. Subtraction of the equation
\begin{eqnarray}
      \Phi^{n}_{j,k,l} =\Phi^{n-1}_{j,k,l} -2 K^{n-\frac{1}{2}}_{j,k,l}\Delta t
\end{eqnarray}
from Eq.~(\ref{eq:lf21}) and elimination of $K$ using Eq. (\ref{eq:lf22})
shows that {\bf LF2} is equivalent to the standard centered
second-order scheme for the second differential order in time form of
the  wave equation (\ref{eq:wave}), in which $\Phi$ lies on
integral time levels and $K$ is not introduced.

\subsubsection{Iterative Crank-Nicholson ({\bf ICN})}

The third evolution algorithm, which we refer to as {\bf ICN}, is a
two-iteration Crank-Nicholson algorithm. For an $N$-iteration Crank-Nicholson
algorithm, the following sequence of operations is executed at each  time-step:

\begin{enumerate}
\item Compute the first order accurate quantities
\begin{eqnarray}
     \stackrel{(0)}{\Phi}\!{}^{n+1}_{j,k,l}
         &=& \Phi^{n}_{j,k,l}-2 K^{n}_{j,k,l}\Delta t \nonumber \\
     \stackrel{(0)}{K}\!{}^{n+1}_{j,k,l}
         &=& K^{n}_{j,k,l}-\frac{1}{2}\nabla^2\Phi^{n}_{j,k,l}\Delta t.
\end{eqnarray}

\item \label{item:sweCNloopstart} Starting with $i=0$,
compute the midlevel values
\begin{eqnarray}
    \stackrel{(i)}{\Phi}\!{}^{n+\frac{1}{2}}_{j,k,l}
        &=& \frac{1}{2} \left\{\Phi^{n}_{j,k,l}+
                 \stackrel{(i)}{\Phi}\!{}^{n+1}_{j,k,l} \right\}
                   \nonumber \\
    \stackrel{(i)}{K}\!{}^{n+\frac{1}{2}}_{j,k,l}
        &=& \frac{1}{2} \left\{ K^{n}_{j,k,l} +
              \stackrel{(i)}{K}\!{}^{n+1}_{j,k,l} \right\}.
\end{eqnarray}

\item Update  using levels $n$ and $n+\frac{1}{2}$,
\begin{eqnarray}
   \stackrel{(i+1)}{\Phi}\!{}^{n+1}_{j,k,l}
       &=& \Phi^{n}_{j,k,l}
         -2\stackrel{(i)}{K}\!{}^{n+\frac{1}{2}}_{j,k,l} \Delta t
                     \nonumber \\
   \stackrel{(i+1)}{K}\!{}^{n+1}_{j,k,l}
       &=& K^{n}_{j,k,l}
        -\frac{1}{2}\nabla^2 \stackrel{(i)}{\Phi}\!{}^{n+\frac{1}{2}}_{j,k,l}
        \Delta t.
\end{eqnarray}
\item Increment $i$ by one and return to step \ref{item:sweCNloopstart}
until $i=N$ is reached.
\end{enumerate}

A stability analysis by Teukolsky shows that the evolution scheme  is
stable for $N=2$ and $N=3$ iterations, unstable for $N=4$ and $N=5$,
stable for $N=6$ and $N=7$, etc.~\cite{teuk}.

\subsubsection{The Courant-Friedrichs-Lewy condition}

The stability of the three evolution algorithms requires they obey the
Courant-Friedrichs-Lewy (CFL) condition that the numerical domain of
dependence contain the analytic domain of dependence, a common
requirement for explicit algorithms. For the staggered leapfrog ({\bf
LF2}) and the iterative Crank-Nicholson ({\bf ICN}) algorithms, we
set $\Delta t=\Delta x/4$ and for the standard leapfrog ({\bf LF1}) we set
$\Delta t=\Delta x/8$, in all cases slightly less than half the CFL
condition for the algorithm.

\subsubsection{Boundary conditions}

Boundary conditions are implemented computationally in the following way,
which we illustrate in terms of scalar wave boundary data specified at
$z=0$ in terms of a function $f$.

The Dirichlet condition
\begin{equation}
     \Phi(t,x,y,0) = f(t,x,y) \label{eq:Dirichlet}
\end{equation}
is straightforward to implement as
\begin{equation}
      \Phi^{n}_{j,k,0} =f^{n}_{j,k} .
\end{equation}
The Neumann condition
\begin{equation}
     \partial_z \Phi(t,x,y,0) = f(t,x,y) \label{eq:Neumann}
\end{equation}
 is implemented as a 3-point
one-sided derivative
\begin{equation}
      \Phi^{n}_{j,k,0} =\frac{1}{3} (-\Phi^{n}_{j,k,2}
                +4\Phi^{n}_{j,k,1}-2 \Delta x f^{n}_{j,k} ) .
\label{eq:gneum}
\end{equation}
The Sommerfeld condition
\begin{equation}
    (\partial_t -\partial_z)\Phi(t,x,y,0) = f(t,x,y), \label{eq:Sommer}
\end{equation}
is implemented in the interpolative form used in several relativity
codes~\cite{bmasso,shibnak,baumgsh} by modeling the field in the
neighborhood of the boundary as $\Phi(t+z,x,y)$ and using a 3-point
spatial interpolation to obtain

\begin{eqnarray}
      \Phi^{n}_{j,k,0} & = &
           \frac{1}{2}\left(2-\frac{\Delta t}{\Delta x}\right)
           \left(1-\frac{\Delta t}{\Delta x}\right)
\Phi^{n-1}_{j,k,0}
         + \frac{\Delta t}{\Delta x}
           \left(2-\frac{\Delta t}{\Delta x}\right)
\Phi^{n-1}_{j,k,1}  \nonumber \\
        & & - \frac{1}{2}\left(1-\frac{\Delta t}{\Delta x}\right)
           \frac{\Delta t}{\Delta x}
\Phi^{n-1}_{j,k,2}
          +\Delta t f^{n-1}_{j,k}  .
\label{eq:intsomm}
\end{eqnarray}

\section{Stage 1: Robust evolution stability}
\label{sec:boun}

Periodic boundary conditions are equivalent to a toroidal topology and
do not introduce the local effects of a real boundary. They provide a test of
the evolution code isolated from the effects of boundary conditions. Because
an instability in such a code may not be evident for a considerable time if
masked by a strong initial signal, the use of random data is efficient at
revealing instabilities early in the evolution. Random initial data does not
satisfy the constraints but that poses no difficulty here, where we are only
concerned with stability. These observations motivate our Stage 1 test bed:

\begin{verse}
{\bf Stage 1}: {\em Run the evolution code on a 3-torus with random
initial Cauchy data. The stage is passed if the Hamiltonian constraint
$C$ does not exhibit exponential growth.}
\end{verse}

An evolution code which does not exhibit exponential growth under these
conditions is defined to be {\bf robustly stable}. Failure at Stage 1 would
rule out applications with boundaries.

We use an evolution time of 2000 crossing times ($2000L$, where $L$ is the
linear size of the computational domain) on a uniform $48^3$ spatial grid with
a time step slightly less than half the Courant-Friedrichs-Lewy limit. These
conditions are computationally practical and are used to determine whether
there is exponential growth of the Hamiltonian, as measured by the
$\ell_\infty$ norm. {\bf All runs reported in this paper are made with these
specifications.}

\subsection{Stage 1 results}
\label{sec:stage1}

We have applied the Stage 1 test to determine whether any of three evolution
algorithms, {\bf LF1}, {\bf LF2} and {\bf ICN} are intrinsically unstable. We
apply the test on the flat 3-torus determined by the periodicity conditions
$h_{ij}(x,y,z)=h_{ij}(x+L,y,z)=h_{ij}(x,y+L,z)=h_{ij}(x,y,z+L)$. The Cauchy
data, $(h_{ij},\dot h_{ij})$, can be initialized as random numbers
in any interval $(-A,A)$, since the system is linear. Here we use the interval
$(-10^{-6},+10^{-6})$.

When applied to the scalar wave equation in the hybrid first order in time and
second order in space form of Eq. (\ref{eq:canwave}), all three algorithms {\bf
LF1}, {\bf LF2} and {\bf ICN} pass Stage 1. This confirms, for the case of
random data, prior work~\cite{sfrg} showing that the hybrid system has a well
behaved computational evolution.

Furthermore, when applied to the ADM system (\ref{eq:ewave}) for gravitational
evolution with $\lambda$ equal to 0, 2 and 4, all three algorithms {\bf LF1},
{\bf LF2} and {\bf ICN} also pass Stage 1. For runs with $\lambda$ equal
to -0.1, 4.1 and 5.0 these three evolution algorithms exhibited exponential
growth. This indicates a range of stability for $0\le\lambda\le 4$.
In this range, it is notable that the norm of the
Hamiltonian constraint grows linearly in time for {\bf LF1} and {\bf LF2} but
decays exponentially for {\bf ICN}. This apparently results from the
artificial dissipation inherent in {\bf ICN}.

The stability of the discretized system of ADM equations is more restrictive
than (but consistent with) the range $-1 \le \lambda$ found by
Frittelli~\cite{frit} for stable evolution of the constraints in the continuum
theory. For $\lambda =-1$, algorithms {\bf LF1} and {\bf LF2} show exponential
growth whereas the norm of the Hamiltonian only grows linearly for {\bf ICN}.
However, for $\lambda=-1.01$ or $\lambda=-0.99$ this norm grows exponentially
for {\bf ICN}. This anomalous behavior suggests that the the special case
$\lambda=-1$ (the Einstein system of evolution equations) can be successfully
evolved but that its numerical stability is highly sensitive to the choice of
finite difference scheme. For that reason, we have not investigated this case
in the presence of a boundary. The upper limit of the window of
stability at $\lambda=4$ is related to the size of the time step. For
algorithm {\bf LF2}, a run with $\Delta t= \Delta x/8$ (half the time step of
the standard runs) and $\lambda =20$ showed no exponential growth. This seems
to arise from the increase of the constraint propagation speed with $\lambda$,
which makes the Courant-Friedrich-Lewy condition more stringent.

In summary, {\bf the hybrid scalar wave system (\ref{eq:canwave}) and
the ADM system (\ref{eq:ewave}), with $\lambda$ equal to 0, 2 and 4,
pass Stage 1 for the three evolution algorithms {\bf LF1}, {\bf LF2} and
{\bf ICN}}. These are the evolution systems whose boundary stability we
investigate in Sec. \ref{sec:stage2}.

\section{Stage 2: Robust boundary stability}
\label{sec:stage2}

The general linear hyperbolic equation in second order differential form for a
scalar field has a well posed Cauchy problem in a region with  Dirichlet,
Neumann or Sommerfeld boundary conditions (e.g., see~\cite{egorov}). For a
system of coupled scalar fields, or a tensor field with coupled components,  it
is standard practice to reduce the equations to first order differential form
in order to examine hyperbolicity and appropriate boundary
conditions~\cite{renfried}. For a first order system  in diagonalizable,
strongly hyperbolic form there is a straightforward way to decide which
variables require data at a given boundary~\cite{kreiss}. Variables propagating
along future directed characteristics which emanate {\it from} the boundary can
be assigned free data, but assigning arbitrary boundary values to the remaining
variables would be inconsistent with the evolution equations. When a second
order system is reduced to first order form, spatial derivatives of the field
become auxiliary variables, so that there is no longer any natural distinction
between, say, Dirichlet and Neumann boundary conditions. What might have been
termed a Neumann condition in the original system now appears in Dirichlet
form. Friedrich and Nagy~\cite{nagy} have recently given a complete treatment
of a well-posed boundary-initial-value problem for a symmetric hyperbolic
version of the nonlinear vacuum gravitational equations. They find a
continuum of allowed boundary conditions on the field variables, which can be
more naturally distinguished as ranging from  ``electric'' to ``magnetic''.

The hybrid form of the scalar wave equation(\ref{eq:canwave}) does not not fit
into any hyperbolic category but, since the spatial derivatives of the field
are not treated as auxiliary variables, we retain the classification of
Dirichlet, Neumann and Sommerfeld boundary conditions. Following common
practice, we also retain this classification in the case of the ADM system
(\ref{eq:ewave}).

Whereas Stage 1 tests stability of the interior evolution algorithm itself,
Stage 2 is designed to be a simple stability test of the combined
evolution-boundary algorithm. The boundary algorithm by itself is neither
stable nor unstable; rather the combination of the boundary algorithm with a
(stable) evolution algorithm may be stable, and a combination with another
(stable) evolution algorithms may be unstable~\cite{strik}.  In Stage 2, the
three torus is opened up in the $z$-direction to form a space of topology
$T^2\times[0,L]$, with boundaries at $z=0$ and $z=L$ coinciding with planes of
grid points. A boundary algorithm for these points is necessary in order to
update the evolution at grid points neighboring the boundary. One purpose of
the testbed is to measure suitability for matching the Cauchy evolution to an
exterior numerically generated solution, such as in CCM, where interpolations
between the exterior and interior grids continually introduce random error at
the Cauchy boundary. This motivates the following criterion for robust
evolution-boundary stability:

\begin{verse} {\bf Stage 2}: {\em Run the evolution-boundary code on
$T^2\times[0,L]$ with random initial Cauchy data and random boundary data. The
stage is passed if the Hamiltonian constraint $\cal C$ does not exhibit
exponential growth.} \end{verse}

As an illustration of how Stage 2 is implemented, rather than giving smooth
Dirichlet data, such as the homogeneous data $\Phi (t,x,y,0)=0$ for a scalar
field, we require that $\Phi$ be prescribed as a random number at each boundary
point. Similarly, in the Neumann or Sommerfeld cases, $\partial_z \Phi$ or
$(\partial_t -\partial_z)\Phi $ are prescribed as random numbers at $z=0$. In
order to avoid inconsistencies, the initial and boundary data are both set to
$0$ in a few grid zones near the intersection of the initial Cauchy surface
with the boundary.

As a first set of experiments, we have confirmed that the hybrid scalar system
(\ref{eq:canwave}) passes Stage 2 for the three evolution algorithms {\bf LF1},
{\bf LF2}and {\bf ICN} with a Dirichlet boundary algorithm. Sommerfeld and
Neumann boundary algorithms were less successful, as indicated in Table 1. Only
those combinations of evolution and boundary algorithms which are robustly
stable for a scalar field should be expected to pass Stage 2 for the ADM
system.

\begin{figure}
\centerline{\epsfxsize=6in\epsfbox{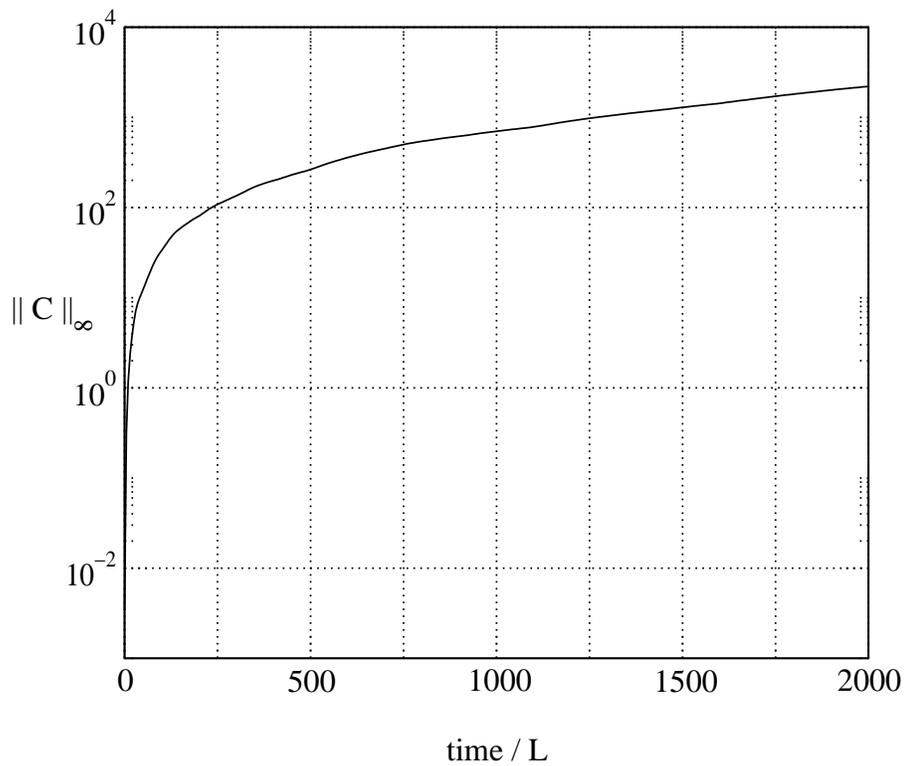}}
\caption{A log plot of the $\ell_\infty$ norm of the Hamiltonian
constraint as a function of crossing time for a Stage 2 test of random
Sommerfeld boundary conditions on all metric components.}
\label{fig:stage2som}
\end{figure}

Next, we tested ADM evolution with boundary data prescribed for each component
of $K_{ij}$, as has been common practice. In Sec.~\ref{sec:st2bc}, for the case
$\lambda =0$, we show this practice leads to an inconsistent evolution-boundary
problem, whose finite difference solutions cannot in general converge to a
correct continuum solution. It is notable that the numerical results
were quite mixed, not necessarily showing unstable growth. For {\em
homogeneous} boundary conditions and evolution with $\lambda=0$, we first found
that the only stable combination was {\bf ICN} evolution with a homogeneous
Sommerfeld boundary. (All other combinations showed exponential growth on the
order of 10 crossing times.)  Next, we applied {\it random} Sommerfeld boundary
data to all components of $K_{ij}$ (the analogue of choosing $f$ randomly in
Eq.~(\ref{eq:Sommer})), again with {\bf ICN} evolution and $\lambda=0$. The log
plot in Fig.~\ref{fig:stage2som} shows the Hamiltonian constraint growing at
late times as $t^n$, for $n\approx 1.92$. Such polynomial growth is normally
regarded as stable. However, in this case, there is a large multiplying
constant, and the magnitude of the error (of the order of 1000 at $t=2000$) is
unacceptably high.

\section{New ADM boundary algorithms}

\label{sec:st2bc}

\subsection{Consistency of ADM boundary conditions}
\label{sec:inconsistent}

Various types of boundary conditions can be applied to a scalar wave, e.g.
Dirichlet, Neumann or Sommerfeld.  There are more options in the ADM case,
corresponding to Dirichlet, Neumann or Sommerfeld conditions on the various
components of the metric, or equivalently on the components of extrinsic
curvature $K_{ij}$.  However, many of these versions are inconsistent
with the evolution or constraint equations. In this regard, we list some
combinations of the linearized Einstein equations and their implications for a
correct boundary algorithm. We take the boundary to be a surface
$z=const$ and denote the transverse directions by $x^A=(x,y)$.

\begin{itemize}

\item The linearized Einstein equation component
\begin{equation}
   2\, {}^{(4)}\tilde G_z^z \equiv 2\dot K^A_A +\partial^B\partial_B h^A_A
             - \partial_A\partial_Bh^{AB} =0,
\label{eq:gzz}
\end{equation}
can be applied on the boundary to
evolve the transverse trace $K^A_A = K_{xx}+K_{yy}$, given the
transverse-tracefree components $K_{AB}-\frac{1}{2}\delta_{AB}K^C_C $.

\item The linearized Ricci tensor equation
\begin{equation}
     {}^{(4)}\tilde R^t_t\equiv {}^{(4)}\tilde R^k_k
                -2{\cal C}\equiv -\dot K =0
\label{eq:rtt}
\end{equation}
can be applied on the boundary to
evolve the trace $K$, thus determining $K_{zz}$ in terms of transverse
components.

\item The Einstein equation components
\begin{equation}
   2\, {}^{(4)}\tilde G_z^A \equiv -2\dot K_z^A -\partial_B(\partial^B
h_z^A-\partial^A h_z^B)
         -\partial_z\partial^A h^B_B +\partial_z\partial^B h^A_B =0 .
\label{eq:gza}
\end{equation}
can be applied on the boundary to
determine $K_z^A$, given the transverse components.

\item The linearized momentum constraint
\begin{equation}
      {\cal C}^A \equiv \partial_z  K^{Az}+\partial_B  K^{AB}
      -\partial^A  K =0
\label{eq:ca}
\end{equation}
or the combination of the time derivative
of the momentum constraints with Eq. (\ref{eq:rtt}),
\begin{equation}
   \dot {\cal C^A} - \partial^A {}^{(4)}\tilde R^t_t
     \equiv \partial_z \dot K_z^A +\partial_B \dot K^{AB} =0,
\label{eq:cadot}
\end{equation}
give other ways to update the Neumann boundary data for $\partial_z
K_z^A$ in terms of Dirichlet boundary values of $K_{AB}$.

\item The combination
\begin{equation}
   \dot {\cal C}^z - \partial^z {}^{(4)}\tilde R^t_t
     \equiv \partial_z \dot K_{zz} +\partial_A \dot K_z^A =0
\label{eq:cz}
\end{equation}
can be used to update the Neumann boundary data for $K_{zz}$.

\end{itemize}

In the symmetric hyperbolic treatment of the Einstein equations
by Friedrich and Nagy~\cite {nagy}, only 2 components of the Weyl tensor
can be prescribed as free boundary data. It would thus be surprising if
free boundary values could be assigned to all metric
variables (or their associated momentum variables) for an ADM system with
gauge freedom fixed by an explicit choice of lapse and shift, as shown by
the following proposition.

{\bf Proposition}: Prescription of Dirichlet boundary data on all components
of the metric (or extrinsic curvature) of the ADM system (\ref{eq:ewave}) with
$\lambda =0$ gives rise to an inconsistent evolution-boundary problem. The same
is true for Neumann or Sommerfeld boundary data.

{\em Proof}: Consider homogeneous Dirichlet data consisting of setting {\it
all} components of $h_{ij}$ to zero on the boundary. Then the function $\Psi:=
\partial_A \partial^A h^B_B-\partial^A\partial^B h_{AB}$ vanishes on the
boundary and  Eq. (\ref{eq:gza}) (one of the evolution equations for this
system) implies that the normal derivative $\partial_z \Psi$ also vanishes on
the boundary. But it is easy to verify, in the case $\lambda =0$, that the {\em
evolution equations} for $h_{ij}$ imply that $\Psi$ satisfies the scalar wave
equation. Thus the vanishing Dirichlet data for $h_{ij}$ generates, for any
initial data, a solution $\Psi$ of the wave equation whose Dirichlet and
Neumann boundary data both vanish. This a classic example of an inconsistent
boundary value problem for the scalar wave $\Psi$. (This is evident from
considering the fate of an initial pulse of compact support when it reaches the
boundary.)

Similarly, consider the homogeneous Sommerfeld data $(\partial_t -\partial_z)
h_{ij}=0$ applied to {\it all} metric components on the plane boundary at
$z=0$. If $h_{ij}$ were a global solution consistent with this boundary data
then, since the equations are linear and have space-time translational
symmetry, $\hat h_{ij}=(\partial_t -\partial_z) h_{ij}$ would also be a global
solution but with vanishing Dirichlet data for all components at the boundary.
Thus, as in the Dirichlet problem, a Sommerfeld boundary condition, or by the
same argument a Neumann boundary condition, applied to all components of the
metric also leads to an inconsistent boundary value problem. $\Box$

\subsection{Robustly stable Dirichlet evolution-boundary algorithms}

In order to formulate consistent boundary algorithms, we denote by $h_{TT}$ the
traceless part of the components transverse to the boundary, i.e.
$(h_{xx}-h_{yy})$ and $h_{xy}$ in our Stage 2 test with boundaries at $z=0$ and
$z=L$. Since our gauge choice $h_{t\mu}=0$ is consistent with the radiation
gauge subclass of harmonic coordinates, these $TT$ components represent the
free modes of waves propagating normal to the boundary. We make the hypothesis
that the boundary values of $h_{TT}$, or equivalently $K_{TT}$, should be
freely specified in either Dirichlet, Neumann or Sommerfeld form. This is
motivated in the Dirichlet case by the consistency of
characteristic evolution where the free data on a worldtube corresponds to
Dirichlet data for $h_{TT}$ in the linearized approximation. Given this $TT$
boundary data, the boundary algorithm must determine boundary values of the
remaining components using the linearized gravitational equations.

{\bf The following five Dirichlet boundary algorithms exhibit Stage 2 robust
stability for the {\bf ICN} evolution algorithm.} The algorithms update the
boundary values of the extrinsic curvature, with boundary values for the metric
perturbation updated by the centered difference version of the first of Eq's
(\ref{eq:ewave}). Given random initial and boundary data for the
transverse-traceless components $K_{TT}$, all five boundary algorithms update
the boundary values of the trace $K^A_A$ via integration of Eq. (\ref{eq:gzz}).
Boundary values of the remaining unspecified components are updated as follows:

\begin{itemize}

\item {\bf \bcfive}: We apply Eq. (\ref{eq:cz}) to update $K_{zz}$ and
Eq.  (\ref{eq:gza}) to update $K_z^A$.

\item {\bcone}: We apply Eq. (\ref{eq:rtt}) to update
$K_{zz}$ and the momentum constraint Eq. (\ref{eq:ca}) to supply
boundary values for $\partial_z K_z^A$ which, expressed as a 3-point
sideways finite difference, are used to update $K_z^A$ .

\item {\bf \bctwo}: We apply Eq. (\ref{eq:rtt}) to update boundary
values for $K_{zz}$ and Eq. (\ref{eq:cadot}) to supply boundary values
for $\partial_z \dot K_z^A$ which, as in \bcone , are used to update
$K_z^A$ using a centered time difference.

\item {\bf \bcfour}: We apply Eq. (\ref{eq:rtt}) to update $K_{zz}$
and Eq. (\ref{eq:gza}) to update $K_z^A$.

\item {\bf \bcthree}: We apply Eq. (\ref{eq:cz}) to update $K_{zz}$
(with finite difference stencils as above) and Eq. (\ref{eq:cadot}) to
update $K_z^A$.

\end{itemize}

\begin{figure}
\centerline{\epsfxsize=6in\epsfbox{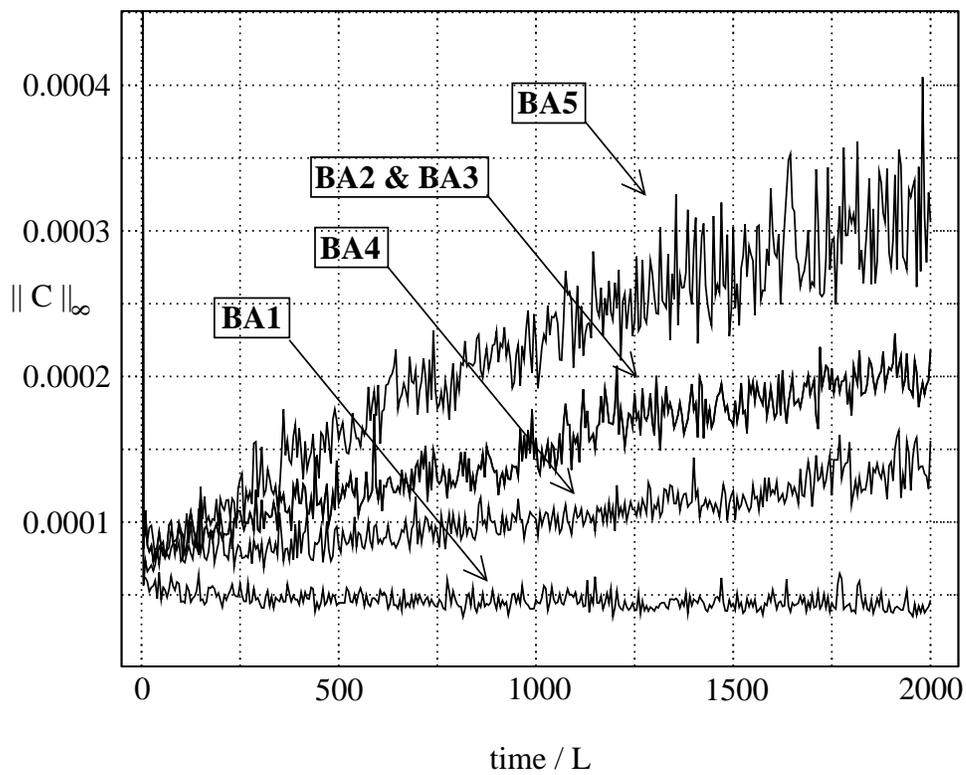}}
\caption{Stage two performance of the Hamiltonian constraint as a function
of crossing time for the five robustly stable algorithms \bcfive\ to
\bcthree\ .}
\label{fig:stage2}
\end{figure}

All five boundary algorithms satisfy Stage 2 robust stability for {\bf ICN}
evolution. Fig. \ref{fig:stage2} shows the behavior of the Hamiltonian
constraint for these five  algorithms in the case $\lambda=2$. Note that
\bcone\
and \bctwo\ have identical performance, as might be expected as they differ
only with respect to details of initialization at the boundary. \bcfive\ and
\bcfour\ show less noise in the Hamiltonian constraint than the others, with
\bcthree\ showing the largest (although still linear) growth. \bcfive\ gave the
best performance, with the Hamiltonian constraint actually decreasing slowly at
late times.

For $\lambda=0$ and $\lambda=4$, boundary algorithms \bcfive\ and \bcone\ are
also robustly stable for ICN evolution. (The other boundary algorithms were
not checked for these cases in order to conserve computing time).

While these 5 Dirichlet boundary algorithms were robust for {\bf ICN}
evolution, they failed Stage 2 for {\bf LF1} and {\bf LF2} evolution with
$\lambda =$ 0, 2 and 4, with the exponential growth rate typically decreasing
with increasing $\lambda$. Table 2 summarizes the performance for $\lambda =2$.
The failure of these evolution-boundary algorithms for leapfrog evolution, but
not for {\bf ICN}, emphasizes the complexity of the finite difference problem
compared to the corresponding analytic problem.

\subsection{Neumann and Sommerfeld  boundaries}

We attempted to modify the Dirichlet boundary algorithms \bcfive\ to
\bcthree\ to obtain stable evolution with Neumann or Sommerfeld boundary
data specified for the extrinsic curvature components $K_{TT}$.
In the Neumann case, assuming all components of the metric have been
determined at time level $N-1$ and the evolution has been applied to
update all components at level $N$ except at the boundary, we express the
Neumann boundary data $\partial_z K_{TT}$ in finite difference form
according to Eq. (\ref{eq:gneum}) to update $K_{TT}$ on the boundary at
level $N$.  This supplies the necessary data to apply the Dirichlet
boundary algorithms to update all remaining components.

Similarly, in the Sommerfeld case, given that the metric has been
determined at level $N-1$ and the evolution has been applied to update
all components at level $N$ except at the boundary, we apply the
interpolative Sommerfeld condition in finite difference form according
to Eq. (\ref{eq:intsomm}) to update $K_{TT}$ on the boundary at level
$N$.  Again this supplies the necessary data to apply the Dirichlet
boundary algorithms to update all remaining components.

We tried an extensive, although not exhaustive, set of combinations of
evolution algorithms, boundary algorithms and values of $\lambda$ with
Sommerfeld or Neumann conditions applied to the $TT$ components but we were
unable to obtain acceptable Stage 2 evolution.

\section{Tests with a cubic boundary (Stage 3)}
\label{sec:stage3}

For application of these algorithms to an isolated astrophysical system,
we next perform tests with a cubic boundary. This is the standard
boundary geometry adopted in Cauchy evolution codes based upon
Cartesian coordinates.  We propose the following
operational criteria of robust stability for a Cartesian
evolution-boundary algorithm for an isolated system:

\begin{verse}
{\bf Stage 3}: {\it Run the evolution-boundary code with a cubic boundary with
random initial Cauchy data and random boundary data. The stage is passed if the
Hamiltonian constraint ${\cal C}$ does not exhibit exponential growth.}
\end{verse}

In view of the Stage 2 results, we confine our Stage 3 investigation to {\bf
ICN} evolution with Dirichlet boundary data on all faces of the cube applied
with the (best performing) boundary algorithm  \bcfive\ .  The edges and
corners of the cube must be handled separately.  The two components $K_{TT} =
-\frac{1}{2} \dot h_{TT}$ are treated as free data (i.e. are specified
randomly) on all faces, edges and  corners. While this means two free
quantities and four update equations on the faces, there are four free
quantities on the edges so that one only needs two update equations.
Furthermore, at any corner, the six $K_{TT}$ components from the neighboring
faces, $K_{xy}$, $K_{xz}$, $K_{yz}$, $K_{xx}-K_{yy}$, $K_{xx}-K_{zz}$ and
$K_{zz}-K_{yy}$, are reduced to five that are independent and therefore freely
specifiable by means of the identity $ [K_{xx}-K_{yy}] + [K_{yy}-K_{zz}] +
[K_{zz}-K_{xx}] = 0$. Thus only one equation is needed to update the corners.

As just indicated, all non-diagonal components are freely specified $TT$
data on the corners. Given the additional $TT$ data $[K_{xx}-K_{yy}]$
and $[K_{zz}-K_{xx}]$, the missing diagonal component $K_{xx}$ is
computed from $$3K_{xx}=K+[K_{xx}-K_{yy}]+[K_{xx}-K_{zz}],$$ where $K$
is updated using the equation $${}^{(4)}\tilde R^t_t =-\dot K=0.$$

It remains to give the algorithm for the edges.  On the edges parallel
to the $x$-axes, $K_{xy}$ and $K_{xz}$ are specified as boundary data.
The missing non-diagonal component $K_{yz}$ is updated using
$^{(4)}\tilde G_{yz}=0$, the same equation used on the neighboring
faces except now the derivatives of the metric in the $y$ and $z$
directions must be computed by 3-point sideways differencing.  The
diagonal components of $K_{ij}$ are computed the same way as on the
corners.

We should note that the routine that solves the constraint
\begin{equation}
   (- \dot {\cal C}^n + \partial^n{}^{(4)}\tilde R^t_t) = 0
\end{equation}
on a face of the cube with normal in the $n$-direction must be called
{\em after} the missing non-diagonal components have been updated on the
edges surrounding that face.  Otherwise, in the case of the
$z=const$ face, when computing the quantity $\partial_y K_{yz}$ on the top
time-level, with centered finite differencing, one might use values of
$K_{yz}$ on the edge parallel to the $x$-axis that were not yet updated.

\begin{figure}
\centerline{\epsfxsize=6in\epsfbox{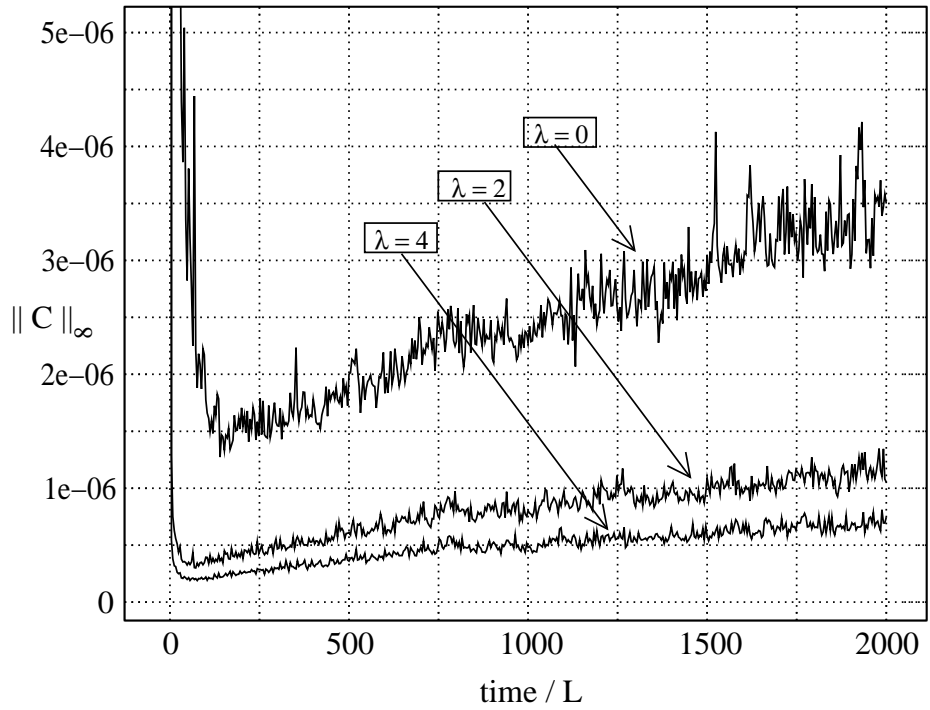}}
\caption{Behavior of the Hamiltonian constraint for a Stage 3 test with
cubic boundary.}
\label{fig:stage3}
\end{figure}

We confirmed that {\bf the above algorithm is robustly stable} by
performing runs with $\lambda = 0, 2, 4$ and
random initial and boundary data. The behavior of the Hamiltonian
constraint as a function of time is shown in Fig. \ref{fig:stage3}.

\section{Conclusion}

We have shown that linearized ADM evolution with boundaries can be carried out
with long term stability in a test bed consisting of random constraint
violating initial data and random boundary data applied to the
trace-free-transverse metric. Adding the Hamiltonian constraint (with $\lambda
>0$) to the Ricci system of linearized equations appears to give better
performance, but does not drastically affect overall robustness. The successful
implementation of an ADM boundary algorithm presented here offers new hope both
for the long term stability of nonlinear ADM evolution and for the prospects of
matching an exterior solution at an ADM boundary. However, this optimism should
be tempered with the following caveats.

Although we have tested our algorithm only for the case of unit lapse and zero
shift, the extension to any explicitly assigned values of the lapse and shift
appears to be straightforward, at least in the linearized theory. However, the
use of dynamical gauge conditions, which couple the values of lapse and shift
to the metric, would require case-by-case reconsideration.

A spherical boundary would be necessary for an application of our algorithms to
CCM, The implementation of a spherical boundary algorithm is simple in
principle. Only the $TT$ metric (or extrinsic curvature) components should be
matched at the boundary, with the the remaining components updated using the
evolution equations. The identification of the $TT$ components can be readily
made in the local tangent space of the boundary. However, a preliminary
investigation reveals nontrivial technical problems arising from the
non-alignment of a spherical  boundary with the Cartesian grid.~\cite{bela}.

Results for the linear theory are important for ruling out approaches that
cannot work in the nonlinear case. However, the real value of our robust
boundary algorithms will depend upon whether they can successfully be applied
to the nonlinear ADM equations. For \bcfive, the best performing boundary
algorithm,  the formal implementation appears to be straightforward (when the
lapse and shift are given explicitly). It is  standard practice in numerical
evolution to choose the boundary to follow the evolution, so that the grid is
propagated up the boundary, This allows straightforward identification of the
$TT$ components of the metric and extrinsic curvature. An examination of the
nonlinear equations used in the boundary algorithm shows that second
derivatives do not appear in any essentially new way that would alter the
finite difference stencils. Nonlinear terms with first time derivatives which
appear in the boundary update scheme can be evaluated either by means of
iterative techniques or in terms of previously known time levels by backwards
differencing. A separate and more problematic issue is the stability of such an
implementation. At the very least, stability in the nonlinear case would
require suppressing the secular modes of the linear theory from becoming
exponential~\cite{aabss}. Preliminary work underway~\cite{neilsen} to
incorporate our boundary algorithms in a nonlinear ADM code shows improved
performance in the weak field regime over applying boundary conditions to all
components of the metric, but it is premature to judge robust nonlinear
stability. Our results for the linearized equations could not have been
obtained without substantial computational experimentation and the same
certainly holds for their extension to the nonlinear case. 

\acknowledgements

We thank B. Schmidt for reading the manuscript and suggesting improvements.
%, as well as the referee for his helpful comments on the presentation of 
% the paper.
We have benefited from conversations with H. Friedrich, S. Frittelli and A.
Rendall. This work has been supported by NSF PHY 9510895, NSF PHY 9800731 and
NSF INT 9515257 to the University of Pittsburgh. N.T.B. thanks the Foundation
for Research Development, South Africa, for financial support, and the
University of Pittsburgh for hospitality. R.G. thanks the
Albert-Einstein-Institut for hospitality. Computer time for this project was
provided by the Pittsburgh Supercomputing Center and by NPACI.

%%%%%%%%%%%%%%%%%%%%%%%%%%%%%%%%%%%%%%%%%%%%%%%%%%%%%%%%%%%%%%%%%%%%%%%%%%%%%%%

%%%%%%%%%%%%%%%%%%%%%%%%%%%%%%%%%%%%%%%%%%%%%%%%%%%%%%%%%%%%%%%%%%%%%%%%%%%%%%%

\begin{table}
\caption{Results of Stage 2 tests for scalar wave evolutions with the {\bf
LF1}, {\bf LF2} and {\bf ICN} algorithms,  using Dirichlet, Sommerfeld and
Neumann boundary conditions on the $z=const$ faces of a cube. A ``$\surd$''
indicates robust stability, a ``$\times$'' indicates exponential instability
and a ``?'' indicates non-linear growth which was not clearly exponential on
the time scale of the test.}

\begin{tabular}{l|ccc}
       & Dirichlet & Sommerfeld & Neumann \\ \hline
LF1    &  $\surd$  &  $\times$  & $?$\\
LF2    &  $\surd$  &  $\surd$   & $?$\\
ICN    &  $\surd$  &  $\surd$   & $?$
\end{tabular}
\label{Table1}
\end{table}

%\bigskip
%\bigskip

\begin{table}
\caption{Stage 2 tests of ADM evolution with $\lambda =2$ for boundary
algorithms {\bf BA1} - {\bf BA5}. A ``$\surd$'' indicates robust stability. A
``$\times$'' indicates instability with the exponential growth rate indicated
in units of crossing time (CT).   }
\begin{tabular}{l|lllll}
        & {\bf BA1} & {\bf BA2} & {\bf BA3} & {\bf BA4} & {\bf BA5}
\\ \hline
LF1     & $\times$ (150 CT)
                    &  $\times$ (300 CT)
                                & $\times$ (300 CT)
                                            & $\times$ (300 CT)
                                                        & $\times$ (300 CT)  \\
LF2     & $\times$ (25 CT)
                    &  $\times$ (300 CT)
                                & $\times$ (100 CT)
                                            & $\times$ (25 CT)
                                                        & $\times$ (100 CT)  \\
ICN     & $\surd $ &  $\surd $  & $\surd $  & $\surd $  & $\surd $
\end{tabular}
\label{Table2}
\end{table}

\end{document}